\documentclass[conference]{IEEEtran}
\IEEEoverridecommandlockouts
\usepackage{cite}
\usepackage{amsmath,amssymb,amsfonts}
\usepackage{algorithmic}
\usepackage{graphicx}
\usepackage{bmpsize}
\usepackage{textcomp}
\usepackage{xcolor}

\usepackage{acronym}
\usepackage{todonotes}
\usepackage{hyperref}
\usepackage{multirow}
\usepackage{svg}
\usepackage{color,soul}
\usepackage{cite}

\usepackage[normalem]{ulem}

\usepackage{subcaption}
\usepackage{siunitx}
\usepackage[mode=buildnew]{standalone}% requires -shell-escape

\definecolor{bworange}{RGB}{230,97,1}
\definecolor{bworangelight}{RGB}{253,184,99}
\definecolor{bwlilalight}{RGB}{178,171,210}
\definecolor{bwlila}{RGB}{94,60,153}

\newcommand{\newJens}[1]{{\color{black}#1}}
\newcommand{\newlan}[1]{{\color{black}#1}}

\newsavebox{\measuredSize}

\def\BibTeX{{\rm B\kern-.05em{\sc i\kern-.025em b}\kern-.08em
		T\kern-.1667em\lower.7ex\hbox{E}\kern-.125emX}}

\begin{document}
	\vspace*{10cm}
	\begin{minipage}{2\linewidth}
		\textcopyright 2022 IEEE. Personal use of this material is permitted. Permission from IEEE must be obtained for all	other uses, in any current or future media, including reprinting/republishing this material for advertising or promotional purposes, creating new collective works, for resale or redistribution to servers or lists, or reuse of any copyrighted component of this work in other works.
		
		This work has been accepted at the 2022 32st International Conference on Field-Programmable
		Logic and Applications (FPL) and will appear in the proceedings and on the IEEE website on/around September, 2022.
	\end{minipage}
	
	\newpage 
	
\acrodef{AES}[AES]{Advanced Encryption Standard}
\acrodef{aes}[AES]{Advanced Encryption Standard}
\acrodef{co}[CO]{Cryptographic Operation}
\acrodef{dut}[DuT]{Device under Test}
\acrodef{dpa}[DPA]{Differential Power Analysis}
\acrodef{sema}[SEMA]{Simple Electromagnetic Analysis}
\acrodef{dema}[DEMA]{Differential Electromagnetic Analysis}
\acrodef{sad}[SAD]{Sum of Absolute Differences}
\acrodef{sca}[SCA]{Side-Channel Analysis}
\acrodef{sdr}[SDR]{Software Defined Radio}
\acrodef{emsca}[EM-SCA]{electromagnetic side-channel analysis}
\acrodef{tra}[TRA]{Template Resynchronization Approach}
\acrodef{tvla}[TVLA]{Test Vector Leakage Assessment}
\acrodef{fft}[FFT]{fast Fourier transform}
\acrodef{snr}[SNR]{Signal to Noise Ratio}
\acrodef{em}[EM]{electromagnetic}
\acrodef{dlla}[DL-LA]{Deep Learning Leakage Assessment}
\acrodef{cpa}[CPA]{Correlation Power Analysis}
\acrodef{sd}[SD]{Standard Deviation}
\acrodef{ots}[OTS]{Off-The-Shelf}
\acrodef{asic}[ASIC]{Application-Specific Integrated Circuit}
\acrodef{fpga}[FPGA]{Field-Programmable Gate Array}
\acrodef{sconr}[SCONR]{Signal of Cryptographic Operations to Noise Ratio}
\acrodef{soc}[SoC]{System-on-Chip}
\acrodef{os}[OS]{Operating System}
\acrodef{lut}[LUT]{Lookup Table}
\acrodefplural{lut}[LUTs]{Lookup Tables}
\acrodef{adc}[ADC]{Analog-to-Digital Converter}
\acrodefplural{adc}[ADCs]{Analog-to-Digital Converters}

\newcommand\avgRoundT{\boldsymbol{a}_{i,w}}
\newcommand\trace{\boldsymbol{t}}
\newcommand\template{\boldsymbol{c}}
\newcommand\templatelength{n}
\newcommand\similarity{s}
\newcommand\offset{o_\mathrm{offset}}
\newcommand\offsetiu{o_{i,\mathrm{upperLimit}}}
\newcommand\offsetil{o_{i,\mathrm{lowerLimit}}}
\newcommand\paral{d}
\newcommand\preci{p}
\newcommand\latency{l}
\newcommand\excesssamples{e_\mathrm{samples}}

\newcommand\datastream{\boldsymbol{T}}

\newcommand\mamo{\texttt{MatcherModule}}
\newcommand\srg{\texttt{SRG}}
\newcommand\comparator{\texttt{Comparator}}
\newcommand\tlogic{\texttt{TriggerLogic}}

\newcommand\colorgt{dashed green}
\newcommand\colorfound{dashed red}

\newcommand\bigO[1]{$\mathcal{O}(#1)$}
	\title{Real-Time Waveform Matching \\  with a Digitizer at 10 GS/s}
	
	\author{\IEEEauthorblockN{Jens Trautmann\IEEEauthorrefmark{1}, Nikolaos Patsiatzis\IEEEauthorrefmark{1}, Andreas Becher\IEEEauthorrefmark{2}, J\"urgen Teich\IEEEauthorrefmark{1}, Stefan Wildermann\IEEEauthorrefmark{1}}
		\IEEEauthorblockA{\IEEEauthorrefmark{1}Friedrich-Alexander-Universit\"at Erlangen-N\"urnberg\\
			Email: \{jens.trautmann,nikolaos.nkp.patsiatzis, juergen.teich, stefan.wildermann\}@fau.de}
		\IEEEauthorblockA{\IEEEauthorrefmark{2}Technische Universit\"at Ilmenau\\
		Email: andreas.becher@tu-ilmenau.de}
		\thanks{This work was supported by Deutsche Forschungsgemeinschaft (DFG, German Research Foundation) as part of the Research and Training Group 2475 "Cybercrime and Forensic Computing" (grant number 393541319/GRK2475/1-2019).}}

	

	\maketitle
	
	\begin{abstract}
\ac{sca} requires the detection of the specific time frame within which \acp{co} take place in the side-channel signal.
In laboratory conditions with full control over the \ac{dut}, dedicated trigger signals can be implemented to indicate the start and end of \acp{co}. 
For real-world scenarios, waveform-matching techniques have been established which compare the side-channel signal with a template of the \ac{co}'s pattern in real time to detect the \ac{co} in the side channel.
State-of-the-art approaches are implemented on \acp{fpga}. 
However, current waveform-matching designs \newlan{process} the samples from \acp{adc} sequentially and can only work with low sampling rates due to the limited clock speed of \acp{fpga}. 
This makes it increasingly difficult to apply existing techniques on modern \acp{dut} that \newlan{operate} with clock speeds in the $\si{\giga\hertz}$ range.

In this paper, we present a parallel waveform-matching architecture that is capable of performing waveform matching at the speed of fast \acp{adc}. 
We implement the proposed architecture in a high-end \ac{fpga}-based digitizer and 
deploy it to detect AES \acp{co} from the side channel of a single-board computer operating at $1~\si{\giga\hertz}$.
Our implementation allows for waveform matching at $10$~GS/s with high accuracy, thus offering a speedup of $50\times$ compared to the fastest state-of-the-art implementation known to us.
\end{abstract}

	\begin{IEEEkeywords}
		Side-Channel Analysis, Hardware Security, Waveform Matching, Parallel FPGA Design, Signal Processing
	\end{IEEEkeywords}

	% \fi
	\IEEEpeerreviewmaketitle
	
	\section{Introduction / Motivation}

Non-invasive physical attacks on implementations are a serious threat to all embedded systems.
Side-channel attacks and fault attacks can break or circumvent cryptographic methods and have been shown to be applicable to many systems (iPhones~\cite{lisovetsLetTakeIt2021}, One-Board Computers~\cite{balaschDPABitslicingMasking2015}, PCs~\cite{lippSoftwarebasedPowerSideChannel2021}, IoT \newlan{(Internet of Things)} devices~\cite{ronenIoTGoesNuclear2017}).
All side-channel-based methods apply triggering mechanisms, which detect the exact time frame \newlan{during which a} specific operation takes place within the side-channel signal, and then activate the data acquisition or the execution of a fault attack.

The most reliable method for triggering is a signal emitted from the device under test which has a constant distance to the operation of interest.
However, in most real-world examples, this is not possible due to:
1) Restrictive access to the device that will not allow a normal user to modify the code, e.g., to generate trigger signals. 
2) Non-deterministic execution due to preemptive scheduling, interrupts, branch prediction, and other methods~\cite{beckersDesignImplementationWaveformMatching2016}.
For example, triggering on I/O \newlan{(Input/Output)} events that can be detected with normal oscilloscope triggering mechanisms such as rising edges, pulses, or level changes are affected by such non-deterministic fluctuations. 
These I/O events \newlan{may then not be equidistant} to the operation of interest for each execution, making it hard to align the different events.

Waveform-matching triggering techniques are state-of-the-art when it comes to circumventing the problems of these approaches. 
A waveform template is used in a pattern-matching system that triggers if the stored template is similar to the waveform of the incoming side-channel signal.
To the best of our knowledge, there arethree different solutions for pattern-based triggering.

All three solutions feature either \ac{sad} or interval matching as a similarity measure at sample rates of approximately $200$~MS/s.
However, many systems currently run at frequencies of over 1GHz. 
Applying \ac{sca} methods like \newlan{\ac{dpa}}~\cite{kocherDifferentialPowerAnalysis1999} or \newlan{\ac{cpa}}~\cite{brierCorrelationPowerAnalysis2004} \newlan{requires} sampling frequencies that are multiple times faster than the frequency of the device under test~\cite{oflynnFrameworkEmbeddedHardware2017}.
All current designs for waveform matching make use of \acp{fpga} to implement sequential pattern matching in real time.
Newer \acp{adc}, which are responsible for the accurate data conversion between the analogue measurement and the digitial acquisition, can already sample at rates well above $1$ GS/s~\cite{buchwaldHighspeedTimeInterleaved2016}.
However, the maximum clock speed of most \ac{fpga} designs is below $1~\si{\giga\hertz}$, which shows that sequential designs will not become fast enough to process at the speed of fast \acp{adc}.

In this paper, we present an \ac{fpga} design that is capable of performing waveform matching at sampling rates of fast \acp{adc} by utilizing a generic parallel architecture that scales with the available number of \acp{lut} in \acp{fpga}. Furthermore, we implement this architecture in a high-end digitizer with a programmable \ac{fpga} and evaluate the resource \newlan{demands} for such a parallel design.
With this digitizer featuring a high memory depth of $4$~GB on-board RAM and a high-speed PCIe connection to the host PC, the implementation offers a powerful one-device solution for \ac{sca} setups in the field.
Further, we can program the \ac{fpga} to record, trigger, and filter traces from any \ac{dut}.
With the pattern matching being implemented within the pipeline of the digitizer, trigger delays \newlan{become inconsequential} for the recording of operations, as the pattern matching happens at line rate.
Furthermore, we show that the design and implementation allows for waveform matching at $10$~GS/s with high accuracy.

	\section{State-of-the-Art}

\subsection{Waveform Matching}
Waveform matching with side-channel information has the goal \newlan{of comparing} a previously obtained pattern, which we will call template $\template$ with length $\templatelength$, to a continuous incoming signal $\datastream$. 
Let $\trace$ be a part of the data stream $\datastream$ at any instance of time.
Trace $\trace$ has the same length $\templatelength$ as the template $\template$. 

If this trace $\trace$ is sufficiently similar to the template $\template$, we can then infer that a specific operation has taken place on the \ac{dut} at that instance of time.
\newlan{Various} methods can be used to calculate such a similarity $\similarity$ that have different qualities and advantages.

\paragraph{Correlation} This similarity measure can be used to find linear dependencies but also to rate the similarity \newlan{between} different time series. 
The Pearson correlation coefficient can be calculated with
\begin{equation}
\similarity_\mathrm{PearsonCorr} = \frac{\sum_{i=1}^\templatelength (\trace_i - \overline{\trace})(\template_i - \overline{\template})}{\sqrt{\sum_{i=1}^\templatelength (\trace_i - \overline{\trace})^2}\sqrt{\sum_{i=1}^\templatelength (\template_i - \overline{\template})^2}} \quad ,
\label{eq:corr}
\end{equation} 
where $\trace_i$ denotes the $i$-th sample of trace $\trace$, and $\overline{\trace}$ is the average over all samples.
The Pearson correlation coefficient can be used to successfully find \acp{co} within a recorded trace and is not influenced by a static vertical offset in the recording~\cite{trautmannSemiAutomaticLocatingCryptographic2022}. The Pearson correlation also corrects for any scaling in the compared time series, which makes it more robust to small variations in the measurement setup.
\newlan{Its} main disadvantage is \newlan{its resource-heavy} implementation, as it uses several multiplications and divisions. 
Additionally, \newlan{it requires} the calculation of several sums and averages over the entire time series, which also makes the correlation computationally too expensive.
This will lead to a long delay, and will use many resources on an \ac{fpga}.
Furthermore, a valid representation of small rational numbers \newlan{must} be chosen and calculations \newlan{must} be made in floating point numbers, which again is resource intensive.

\paragraph{\ac{sad}} As shown in Eq.~\eqref{eq:sad}, the \acf{sad} first calculates a sample-wise absolute difference between the incoming time series and the template, and then sums the results. 
SAD returns a low score if the two series are similar. 
\begin{equation}
\similarity_\mathrm{SAD} = \sum_{i=1}^\templatelength \vert \trace_i-\template_i \vert
\label{eq:sad}
\end{equation} 
The calculation of the \ac{sad} is significantly less resource intensive than the Pearson correlation and is a commonly used similarity indicator for normalized traces. 
However, one disadvantage is the dependency for equal vertical offsets and scaling. 
If these two properties are not equal for the template and the incoming signal, proper detection cannot be guaranteed\newlan{, thereby requiring adjustments to the threshold.} 
Furthermore, single outliers can distort the entire matching process. 
Multiple outliers or measurement errors can lead to an extremely large sum which, in turn, could lead to an overflow, making the implementation less stable.
All summed values must have at least the same resolution as the template and the trace, which implies a complex addition.

\begin{figure}
	\centering
	\includegraphics[width=1\linewidth]{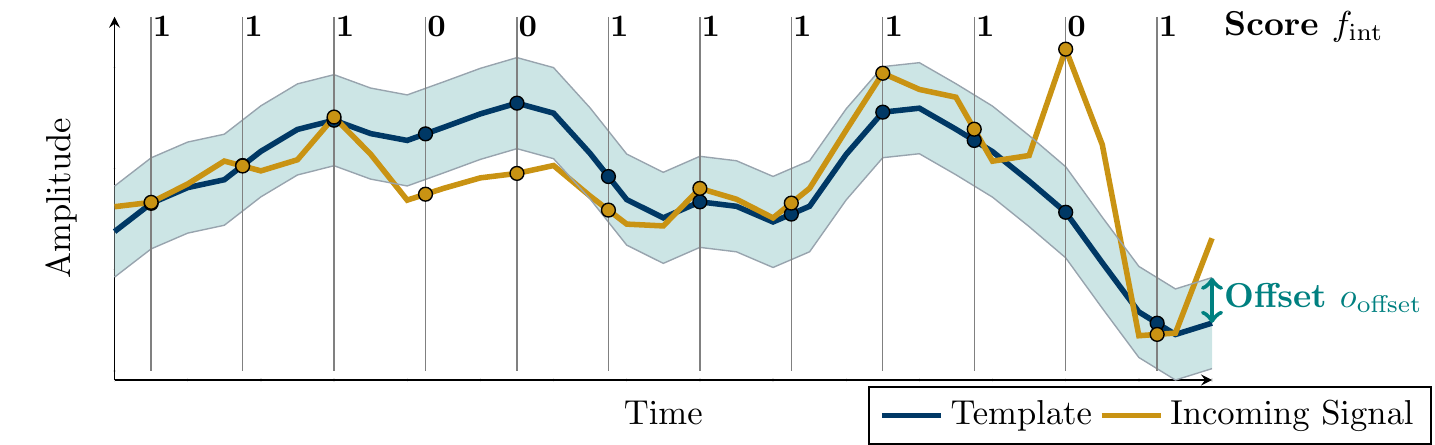}
	\caption{Visual example of waveform matching, which sums the number of sample positions for which the incoming signal (yellow) lies within a corridor around the template (blue line).} 
	\label{fig:intervalmatching}
\end{figure}

\paragraph{Interval Matching} This similarity measure specifies an interval with a chosen offset above and below the established template for every sample position $i$, see Eq.~\eqref{eq:finterval_easy}. 
The similarity is measured as the sum of samples that lie within this interval, see Eq.~\eqref{eq:finterval_sum}. 
A visual explanation is given in Fig.~\ref{fig:intervalmatching}.

\begin{equation}
f_\mathrm{int}(\trace,\template,i,\offset) = 
\begin{cases}
1, & \text{if}\ \template_i + \offset \geq \trace_i \geq \template_i - \offset \\
0, & \text{otherwise.}
\end{cases}
\label{eq:finterval_easy}
\end{equation}
\begin{equation}
	\similarity_\mathrm{Interval} = \sum_{i=1}^{\templatelength} f_\mathrm{int}(\trace,\template,i, \offset)
\label{eq:finterval_sum}
\end{equation}
Similar to \ac{sad}, vertical offsets and scaling are a challenge when using the interval matching method.
\newlan{Success here strongly depends} on the correct calibration of the offset $\offset$ and the matching threshold in $\similarity_\mathrm{Interval}$ when a trace $\trace$ is considered a match.
However, interval matching has the advantage \newlan{of summing binary information alone.}
This leads to simple additions and a maximal sum that is already known at design time and without being arbitrarily large as in \ac{sad}. 
Furthermore, extreme outliers will not have a disproportionate effect on the overall result.
Additionally, with Eq.~\eqref{eq:finterval_easy} we can set the offset $\offset$ and the threshold for the similarity $\similarity_\mathrm{Interval}$.
This provides one additional degree of freedom compared to the other methods, \newlan{which may facilitate a more} accurate calibration.
Due to \newlan{these myriad} advantages, we use interval matching in our implementation.

\subsection{Devices that use waveform matching for triggering}

There are already several solutions that utilize waveform-matching triggering systems. 
All utilize \acp{fpga} as target platforms.
Both icWaves from Riscure \cite{IcWavesSecurityTest} and the ChipWhisperer Pro (CW1200) \cite{CW1200ChipWhispererProNewAE} are two commercial products that use \ac{sad} as a similarity measure.
icWaves is capable of matching an incoming signal to a template at $200$~MS/s with an 8-bit resolution, and the ChipWhisperer Pro can achieve $105$~MS/s with a 10-bit resolution. 
The size $\templatelength$ of the template can, at maximum, be 1024 samples for icWaves and 128 samples for the ChipWhisperer Pro.
The trigger box from Beckers et al.~\cite{beckersDesignImplementationWaveformMatching2016} utilizes interval matching and can match waveforms at $125$~MS/s with a 14-bit resolution.
The maximum length of a template is $1500$ samples, which can be improved if the resolution is lowered.
The trigger box~\cite{beckersDesignImplementationWaveformMatching2016} is constructed with fault injections in mind, which is why its design is heavily tuned to have a low response time \newlan{between the pattern occurring in the side channel and the emission of a trigger signal. } 
In principle, the low sampling rates are not detrimental as long as the side-channel waveform patterns can also be successfully detected with undersampling a faster \ac{dut}'s side channel. 
The waveform matcher can emit a trigger signal to start recording with a higher-end oscilloscope for high-resolution measurements. 
\newlan{This does, however,} increase the complexity of the measurement setup overall. 
Note that the waveform matching would \newlan{still operate with a signal at }a low sampling rate and this might not work if the relevant operation contains high-frequency side-channel patterns.

\newlan{To summarize, state-of-the-art techniques for waveform-template matching often feature sampling rates below the rate necessary for a direct side-channel attack.
Hence, these techniques have to utilize the help of an additional oscilloscope to collect traces from a system operating in the $\si{\giga\hertz}$ range.}
Furthermore, patterns that arise from operations only lasting a few clock cycles might not be detectable at all.
As a remedy, we propose a highly parallel design which can be scaled to even higher sampling rates than $10$~GS/s given the resources on the FPGA and a sufficiently fast ADC.

	\section{Hardware Architecture for Parallel Waveform Matching}\label{sec:hwdesign}

In order to have a significantly higher sample rate of $10$~GS/s and to be able to match waveforms at such high speeds, a parallel architecture is inevitable as most FPGAs will not run faster than $500~\si{\mega\hertz}$ with a sufficiently complex design.
Therefore, we must process multiple samples in each clock cycle in order for the architecture to be scalable.
We denote this degree of parallelism as $\paral$ and need to process $\paral$ samples each clock cycle.
With a higher degree of parallelism, the design can scale with the speed of modern \acp{adc}\newlan{, but must also, at the same time, account for }higher resource requirements in terms of \acp{lut}.
To achieve a design capable of matching waveforms at $10$~GS/s, a \newlan{potential} degree of parallelism where $\paral = 32$ on an \ac{fpga} clocked at $312.5~\si{\mega\hertz}$ would be required, which is an achievable clock speed for modern \acp{fpga}.

\begin{figure*}
	\centering
	\includegraphics[width=0.95\linewidth]{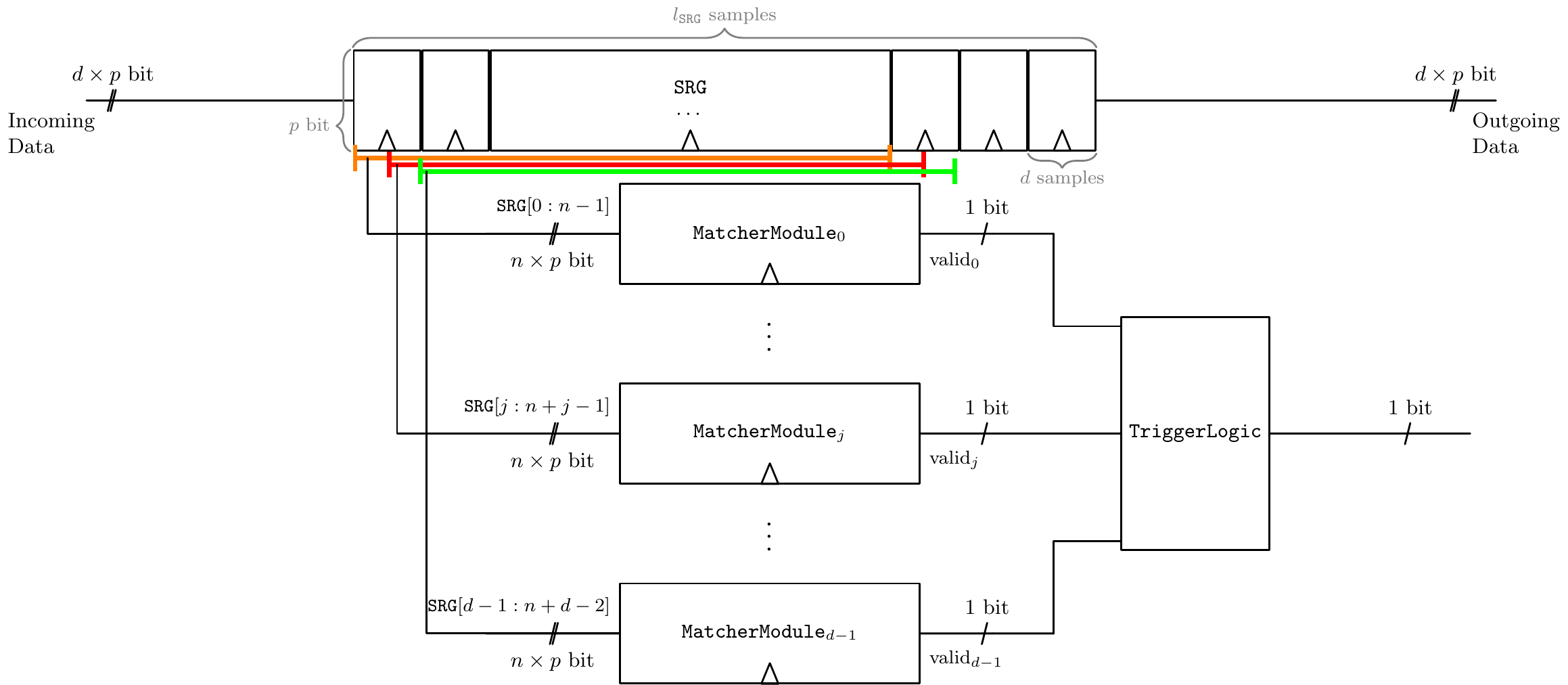}
	\caption{Parallel matching architecture with the \srg{} that takes $\paral$ samples with $\preci$-bit length every clock cycle.}
	\label{fig:schemoverview}
\end{figure*}

Fig.~\ref{fig:schemoverview} \newlan{shows our proposed} the parallel matching architecture.
Our waveform matching module receives $\paral$ samples per clock cycle each with $\preci$-bits of resolution as the input, 
and also outputs $\paral$ samples per clock cycle. 
The memory for buffering samples is implemented as a shift register \srg{} 
of length $l_\text{\srg}$. 
It contains $\paral$ samples at each register stage. 
With each clock cycle, the $\paral$ samples are shifted to the next stage while the new samples are inserted into the first register stage.
As each shift register stage holds $\paral$ samples of $\preci$-bit sizes, they have a width of $\paral \times \preci$ bits.

The length $l_\text{\srg}$ of the shift register is determined \newlan{as follows:} %based on the following criteria.
In order to compare the incoming waveform to the entire template $\template$ of length $\templatelength$, we store at least $\templatelength$ samples.
In practice, more than $\templatelength$ samples should be stored in the \srg{} to first compensate for the latency $\latency$ of the circuit until a match is detected and a valid signal is emitted. 
Second, there might be cases where the template only covers the waveform at the end of the operation (starting at position $l_\mathrm{positionalBuffer}$ within the operation).
However, if the waveform of the entire operation is required for further analysis, all samples of the operation preceding the template should also be stored and transferred to the host PC.
We denote this number of excess samples as $\excesssamples$ calculated as follows:
\begin{equation}\label{eq:excess_samples}
	\excesssamples = \latency \cdot \paral + l_\mathrm{positionalBuffer} .
\end{equation} 
The shift register therefore has a length of 
\begin{equation}\label{eq:length_srg}
	l_\text{\srg} = \templatelength + \excesssamples
\end{equation} 
and a width of at least $\preci$-bit.

As each of the simultaneously obtained $\paral$ samples could be a potential starting position of a \ac{co}, 
segments starting at each of the $\paral$ positions must be compared to the template $\template$. 
Hence, to maintain the throughput, we compare $\paral$ vectors of length $\templatelength$ to the template at each clock cycle. 
This is implemented by $\paral$ matcher modules (see Fig.~\ref{fig:schemoverview}) that all read from their respective part of the shift register. 
Each \mamo$_j$ with $ 0 \le j < \paral$ receives the vector \srg$[j:j+\templatelength]$.
Every \mamo{} must only compare its share of the \srg{} to the template and yield a \texttt{valid} signal which as an indication of a matched sequence starting at position $j$. 
Each of the $\paral$ valid signals is then routed into the \tlogic{} module.
This module ensures that the trigger signal is maintained as a logic \texttt{1} for the duration of the entire operation of interest if one of the \mamo s is emitting a valid signal.
Furthermore, it also acts as a hold-off counter, preventing a trigger signal twice in a row if there are multiple matches within one operation of interest. 
Since the latency of the design is included as per equation ~\eqref{eq:excess_samples} and \eqref{eq:length_srg}, the module thereby synchronizes the emitted trigger signal with the sample output.

\begin{figure}
	\centering
	\includegraphics[width=1\linewidth]{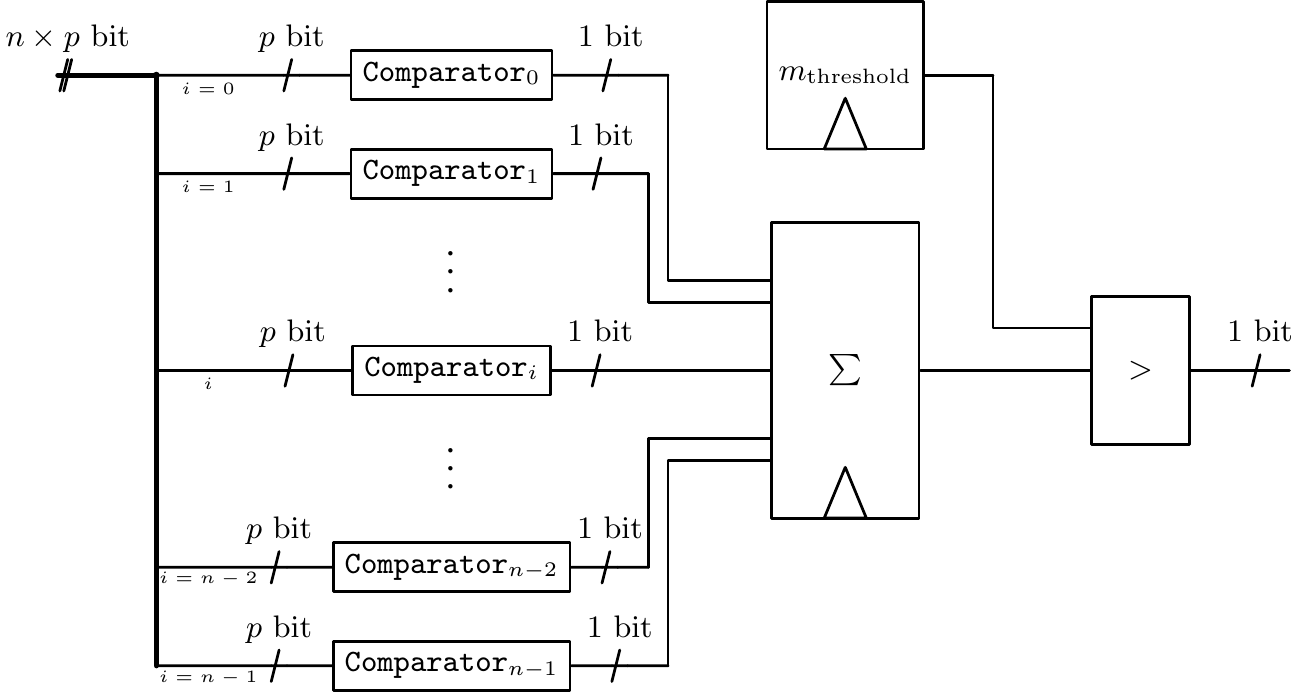}
	\caption{The \mamo{} with all parallel comparators.}
	\label{fig:prelimschematicmatcher}
\end{figure}

A matcher module (see Fig.~\ref{fig:prelimschematicmatcher}) compares the $\templatelength$ samples with the intervals in parallel, sums up the matches, and compares the sum to a threshold. 
With parallelism in mind, we cannot use the same architecture for the \mamo s as the authors of~\cite{beckersDesignImplementationWaveformMatching2016}. 
\newlan{There, Beckers et al.\ compared} the incoming sample from the \ac{adc} to each possible point of the template and used a register adder chain to add up potential matches.
In doing so, they achieved a low latency of only $4$ clock cycles when a pattern similar to the template $\template$ occurs on the incoming stream.
However, we cannot utilize a similar design, since we have to compare multiple samples per clock cycle simultaneously which will not work with a sequential adder chain.
Instead, we use an adder tree to add up the values of the different \comparator{} modules in parallel, displayed as the $\sum$ block in Fig.~\ref{fig:prelimschematicmatcher}.
As we are only adding single bit binary values, we can utilize LUTs for this. 

\begin{figure}
	\centering
	\includegraphics[width=0.7\linewidth]{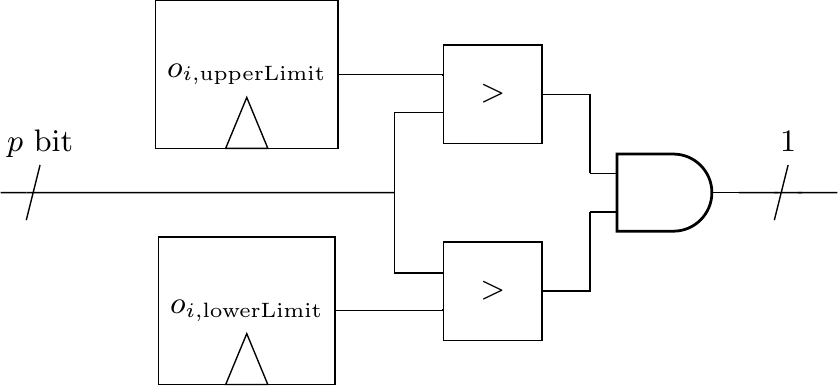}
	\caption{The \comparator$_i$ that is used for each sample $i$ in the template.}
	\label{fig:comparator}
\end{figure}

As we chose interval matching for our design, we compare samples against the upper and lower boundaries. 
The \comparator{} modules (Fig.~\ref{fig:comparator}) are nearly identical to the design in \cite{beckersDesignImplementationWaveformMatching2016}. 
We need to compare each sample $\trace_i$ in the trace $\trace = \text{\srg}[j:j+\templatelength-1]$ to the respective boundaries of the template $\template$ at the same position.
Here, we can directly apply the function $f_\mathrm{int}(\trace,\template,i,\offset)$ from \eqref{eq:finterval_easy} for each sample.
The boundaries are already precomputed for each sample $i$ and substituted with the registers $\offsetiu$ and $\offsetil$.
Also, other similarities measures could be implemented in the \mamo, as long as  the FPGA has sufficient resources to implement the design.

	\section{Implementation of Parallel Waveform Matching at 10GS/s on a Digitizer}\label{sec:impl}

This section discusses the implementation of the proposed parallel architecture within the FPGA of a digitizer.
% 1. Aufbau Digitizer (ADC, Logic, on-board Memory, PCIe)
A digitizer is an electronic acquisition device that \newlan{first} captures analog signals, \newlan{then} samples and digitizes them using \acp{adc}, and finally sends them to a buffer where they can be stored before being processed by a host computer. 
FPGA-based digitizers also allow implementing additional application-specific logic for processing the digitized signal.
%
% 2. Vorteil Digitizer
FPGA-based digitizers thus couple the high sampling rates of modern \acp{adc} directly with the programmability and high parallelism of modern \acp{fpga}.
They thus pose the ideal single-device target platform to implement the proposed parallel waveform matching.

% 3. Eckdaten des verwendeten Digitizers
We use the ADQ7 from SP-Devices~\cite{ADQ7DC10GSPS} which features an onboard Xilinx Kintex UltraScale FPGA (KU85).
It uses two parallel \acp{adc} for recording at $10$~GS/s with a 14-bit resolution.
With $4$~GB of onboard memory, very long traces can be recorded and stored. 
A PCIe interface enables the communication between the digitizer and the host PC. 
As PCIe is limited to $6.8$ GBytes/s and the ADC has a data rate of up to $20$ GBytes/s, a reduction of the data is necessary to continuously monitor a device for \ac{sca}. 
Consequently, for long \ac{sca} measurements, triggering to only record operations of interest (\acp{co}) is essential. 

\begin{figure}
	\centering
	\includegraphics[width=1\linewidth]{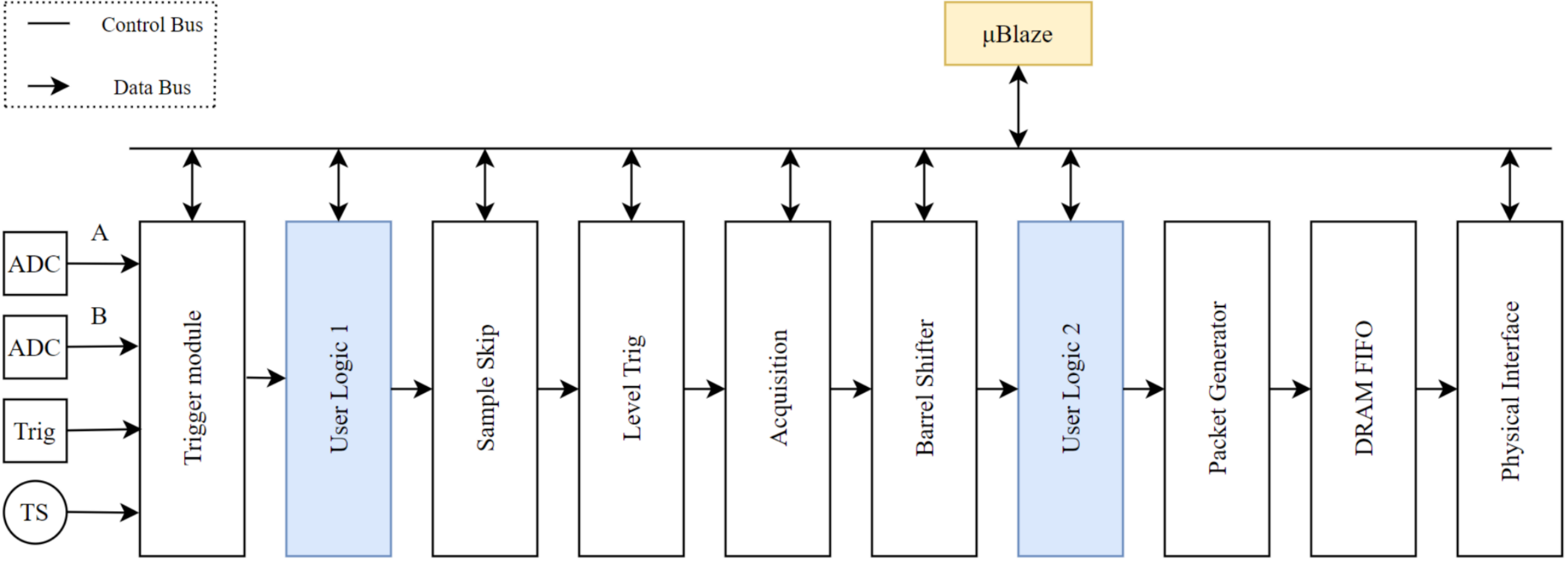}
	\caption{Block diagram of the data path of the digitizer ADQ7DC, adapted from~\cite{ADQ7DCDevelopmentKit}.}
	\label{fig:datapath_adq}
\end{figure}

The datapath of the ADQ design is shown in Fig.\ref{fig:datapath_adq} and has two blocks (shown in blue) representing \textit{User Logic 1} and \textit{User Logic 2}.
Within these blocks, filters and other real-time signal processing can be implemented.
The FPGA is clocked at $312.5~\si{\mega\hertz}$ and has to process $\paral = 32$ samples with $\preci = 14$ bit precision per clock cycle in order to process 10 GS/s.
The parallel matching design presented in Section~\ref{sec:hwdesign} is implemented in \textit{User Logic 1}.

\subsection{Implementation Details}
 \newlan{The following outlines }our \ac{fpga} implementation on the Xilinx Kintex UltraScale FPGA of the digitizer. 
 It employs 6-input \acp{lut} (LUT6) and 8-bit carry logic per slice (CARRY8)\footnote{
 While the presented numbers of \ac{lut} utilization will not exactly match for any implementation on \ac{fpga} with older logic (e.g. LUT4 or CARRY4), the bigger picture will remain the same. 
}.
 
\paragraph{Shift Register} 
The implementation of the shift register \srg{} consists of a cascade of the 32-bit shift register (SRL32) elements found in Xilinx FPGAs.
This is a scalable implementation for shift registers based on the distributed RAM cells in the SLICEM slices, compared to the dedicated flip-flop-based shift register implementation.
Because of the FPGA featuring more than enough LUTRAM, which is the main resource used by the shift register, the shift register could hold up to 30,000 samples.
Our design does not use other primitives like dedicated (Block)RAM and DSPs.
Thus, \acp{lut} are the limiting resource for our design.

\paragraph{Comparators}
These are implemented in \acp{lut}. 
Usually, dedicated carry logic of the Configurable Logic Block (CLB) can achieve a much better performance compared to \acp{lut} for small arithmetic operations thanks to its structure and independent routing resources. 
However, even a multi-level \ac{lut} implementation satisfies the timing requirements \newlan{as our case requires} only a simple comparison with a limited bit width.
If carry-lookahead gates are used, then each comparator module requires 22 \acp{lut} per sample on average. 
On the other hand, for a \ac{lut}-based implementation, only 6 \acp{lut} are needed on a per-sample basis. 
These 6 \acp{lut} alone allow implementing both comparisons contained in the module as well as the conjunction (AND gate) operation of the comparator.
\newlan{Therefore, a \ac{lut}-based implementation, as opposed to the default implementation, is preferred} as it leads to approximately 3.5 times less \ac{lut} usage.

\subsection{Overall Resource Footprint and Comparison}

Waveform matching is implemented in \textit{User Logic 1} of the digitizer's  datapath pipeline shown in Fig.~\ref{fig:datapath_adq}. 
For the implementation, we can thus use the logic that is not yet occupied by other components of the datapath. 
Table~\ref{tab:resources} shows the \ac{lut} usage of our waveform matching module only.
We show the number of \acp{lut} used for the implementation of the comparators utilizing carry logic or following an \ac{lut}-based approach.
The utilization of \acp{lut} in percent is calculated in accordance with the Xilinx KU85 used in the ADQ7DC~\cite{ADQ7DCDevelopmentKit}.
As shown, implementing the adders of the design using the dedicated carry logic would result in \newlan{unnecessarily large} designs. 
By implementing the adders in \acp{lut}, we can implement matchers that scale better:
\begin{table}
	\centering
	\begin{tabular}{|r|r|r|r|r|}
		\hline 
		Template length $\templatelength$ & \multicolumn{2}{c|}{Carry logic} & \multicolumn{2}{c|}{\ac{lut}-based}\\ 
		(samples)& \acp{lut} & \% util. & \acp{lut} &  \% util. \\ 
		\hline 
		2800 & $2,447,523$ & $492\%$ & $680,472$ & $137\%$ \\ 
		\hline 
		1400 & $1,222,969$ & $246\%$ & $338,948$ &  $68\%$\\ 
		\hline 
		700 & $611,651$ & $123\%$ & $169,386$ &  $34\%$\\ 
		\hline 
	\end{tabular} 	
\caption{Resource footprint of the entire parallel waveform-matching implementation using carry logic or a LUT-based adder for the 1-bit additions.}
\label{tab:resources}
\end{table}
For each sample of the template $\template$, we need approximately $244$ \acp{lut}.
Therefore, the approach scales linearly with the number of samples in our template, making it rather easy to fit a specific template into our design.

Table~\ref{tab:comparisonsota} compares our implementation to the trigger box by Beckers et al.\cite{beckersDesignImplementationWaveformMatching2016} and the icWaves solution~\cite{IcWavesSecurityTest}.
With a maximum of $\templatelength = 1,400$ samples as the template length, our implementation is comparable to the state of the art in terms of template lengths, while achieving a much higher sample rate.
With $\templatelength = 1,400$, we need approx.\ $68\%$ of the available \acp{lut}, which leaves resources for other functions being implemented on the \ac{fpga}.
A further reduction of the template length will therefore linearly reduce the needed number of \acp{lut}.
The memory of the ADQ7DC can store $4$~GB of data, making it theoretically possible to store $1,000,000$ k samples with a resolution of 16 bit.
However, for our tests, we never stored more than $100,000$ k samples, which is
sufficient \newlan{for evaluating} the workflow to perform \ac{sca} based on waveform matching.
\begin{table*}
	\centering
	\begin{tabular}{cccc}
		\hline
		 &\textbf{ Our implementation} & \textbf{trigger box}~\cite{beckersDesignImplementationWaveformMatching2016}& \textbf{icWaves}~\cite{IcWavesSecurityTest} \\ 
		\hline 
		\textbf{Algorithm} & Interval matching & Interval matching & SAD  \\ 
%		\hline 
		\textbf{Sample rate} & 10 GS/s & 125 MS/s & 200 MS/s  \\ 
%		\hline 
		\textbf{Max. resolution} & 14-bit & 14-bit & 8-bit  \\ 
%		\hline 
		\textbf{Template length} & $1400$ & $1500$ & $(1 \times 512)$ or $(2 \times 256)$\\ 
%		\hline 
		\textbf{Memory depth} & $100~000$ k & $60$ k & $8~000$ k  \\ 
		\hline 
	\end{tabular} 	
	\caption{Features of our implementation, the trigger box~\cite{beckersDesignImplementationWaveformMatching2016}, and icWaves~\cite{IcWavesSecurityTest}. Values adapted from similar list in~\cite{beckersDesignImplementationWaveformMatching2016}}
	\label{tab:comparisonsota}
\end{table*}

	\section{Workflow and Evaluation} 
We evaluate the proposed technique by performing \ac{sca} based on waveform matching on a BeagleBone Black.
The BeagleBone Black runs Linux based on Debian 10 with kernel version \texttt{4.19.94-ti-r42}, with a TI AM335x ARM processor clocked at a fixed frequency of $1~\si{\giga\hertz}$~\cite{BeagleBoardOrgBlacka}.
The measurement setup is based on the setup proposed by Balasch et al.~\cite{balaschDPABitslicingMasking2015} and features a Langer magnetic near-field probe, model RF-B, on the decoupling capacitor C66 on the back of the BeagleBone.
The \newlan{\ac{em}} signal was amplified by a wideband $30~\si{\decibel}$ amplifier from Langer, model PA303.
We recorded the side-channel signal with the ADQ7DC at $10$~GS/s with a resolution of 14 bits.

The operation of interest was AES-128 encryptions which were performed as part of OpenSSL (\texttt{1.1.1d}) operations. 
We applied OpenSSL on 256 blocks of random data. 
We used the standard OpenSSL encryption without special hardware acceleration or similar optimizations. 
\begin{figure}
	\centering
	\includegraphics[width=1\linewidth]{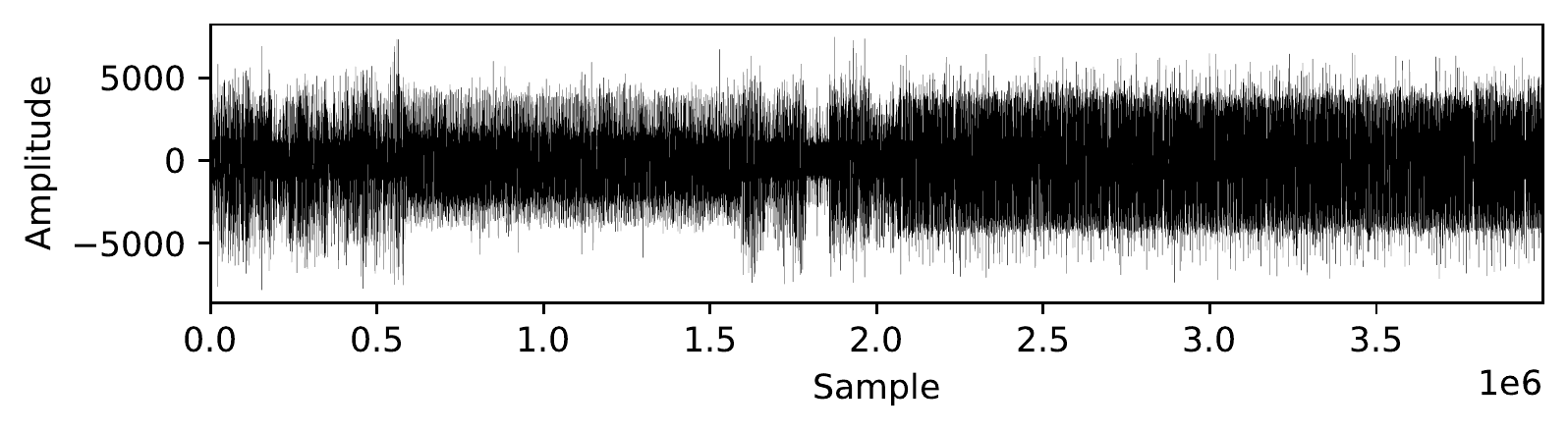}
	\caption{Trace with $256$ AES-128 encryptions performed by the BeagleBone at $1~\si{\giga\hertz}$.}
	\label{fig:beagleboneopensslaes_trace}
\end{figure}

\ac{sca} based on waveform template matching can be divided into three phases (see \cite{schlumbergerCORSICAFrameworkConducting2021}): 1) the calibration stage, 2) measuring stage, and 3) evaluation stage. 
In the calibration stage, a waveform template is constructed, e.g., based on side-channel measurements from the \ac{dut}. 
In the measuring stage, the waveform template is applied for triggering, so that single \acp{co} can be detected on the side channel with pattern matching in real time and recorded for subsequent analysis. 
Finally, in the evaluation stage, the recorded traces of \acp{co} are used for further attacks such as key recovery. 
In the following, we concentrate on phases 1 and 2 which are directly related to waveform matching.

\begin{figure}
	\centering
	\includegraphics[width=1\linewidth]{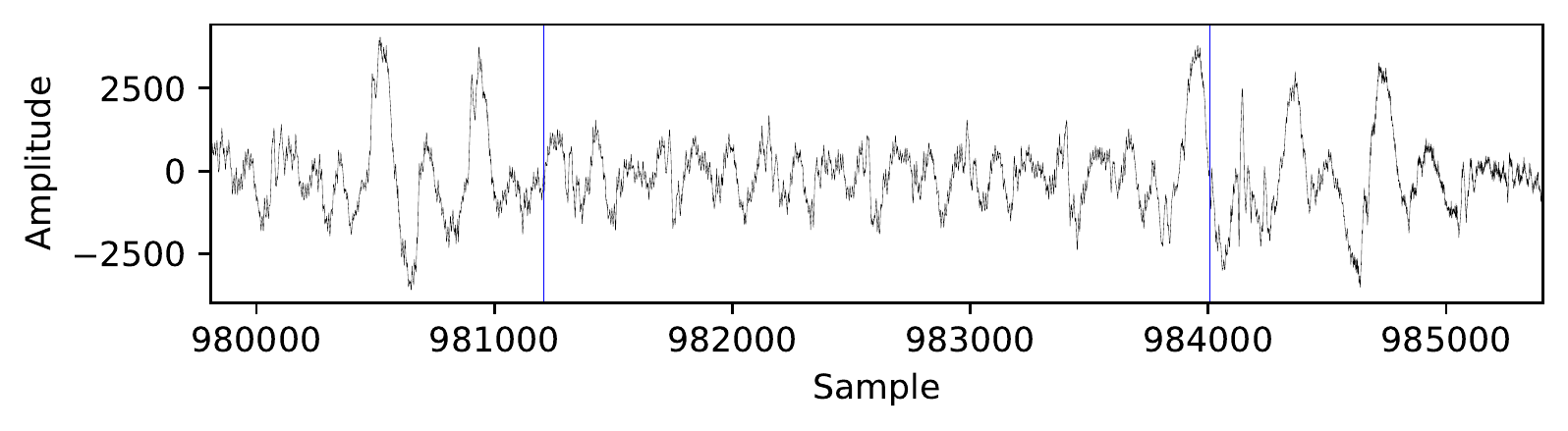}
	\caption{Part of a large trace with one visible AES-128 encryption performed by the BeagleBone at $1~\si{\giga\hertz}$. The vertical blue lines are the start and end of an AES-128 Operation.}
	\label{fig:beagleboneopensslaes_co}
\end{figure}

In order to calibrate our waveform matching design for this experiment, we first recorded the entire encryption process comprising all 256 AES-128 operations without knowing the exact timing of the \acp{co}.
With the help of an algorithm by Trautmann et al.~\cite{trautmannSemiAutomaticLocatingCryptographic2022} for semi-automatically locating the \acp{co} in this trace and the knowledge that one encryption round of AES-128 takes $28$ cycles, we were able to find the locations of the \acp{co}.
The entire trace with all $256$ encryption operations can be seen in Fig.\ref{fig:beagleboneopensslaes_trace}. 
$254$ of the $256$ AES-128 encryptions were successfully found with the semi-automatic approach.
The first encryption operations recorded in this trace are deformed\newlan{, as is always the case for this device, which is} caused by empty caches leading to a different pattern in the side-channel trace~\cite{beckersDesignImplementationWaveformMatching2016}.
The trace of one of the $254$ detected encryption \acp{co} can be seen in Fig.~\ref{fig:beagleboneopensslaes_co}. 

\begin{figure}
	\centering
	\includegraphics[width=1\linewidth]{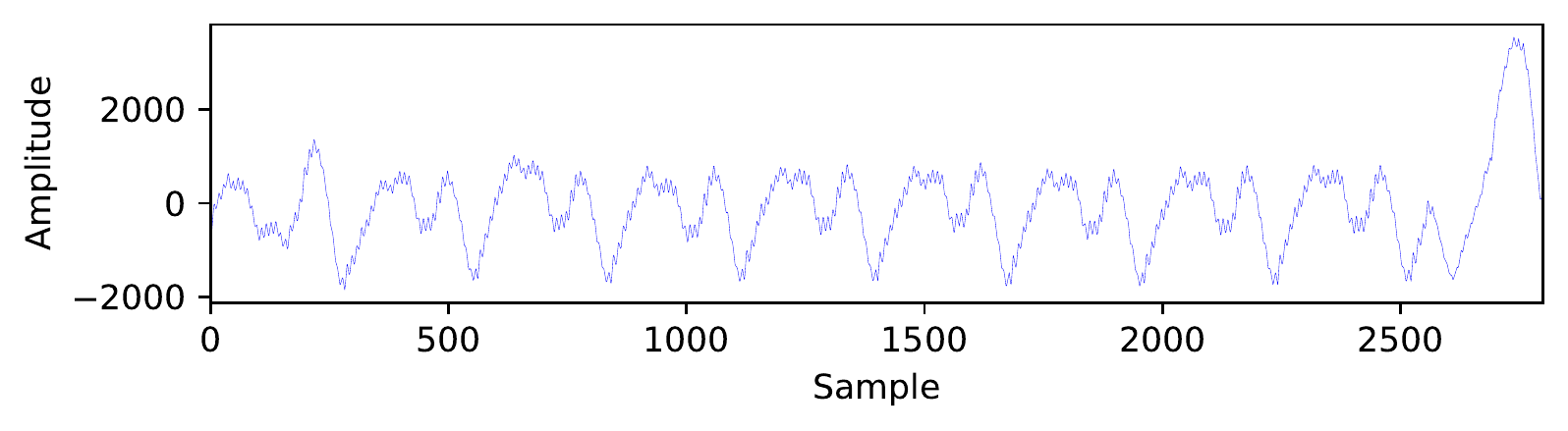}
	\caption{Template for one AES-128 encryption performed by the BeagleBone.}
	\label{fig:beagleboneopensslaes_template}
\end{figure}

%%%%%%%%%%%%%%%%%%%%%%%%%%%%%%%%%%%
% Template generation
An averaged trace of the \acp{co} was then calculated with the information on the locations of the $254$ well-formed \acp{co}. 
The averaged trace is shown in Fig.~\ref{fig:beagleboneopensslaes_template}. 
It can then be used as a template $\template$ for the interval-matching approach.

However, in full resolution, the template has $2800$ samples, which would overwhelm our implementation, as we can only fit $1400$ samples into the current implementation.
A simple solution to this problem is to reduce the number of comparisons between the template and the incoming trace $\trace$ done by the waveform matcher:
It is possible to only compare every $4$-th value in the incoming trace $\trace$ to our template $\template$ and still find the same matches as with a template at full resolution.
While this reduces the number of \comparator{} instances per \mamo{} by a factor of $4$, we still record and acquire data at $10$~GS/s and also \newlan{maintain} the shift register \srg{} at full length.
This is possible because \srg{} only uses LUTRAM, which is not the limiting resource. This stands in contrast to \acp{lut}, which are the bottleneck of this implementation.
%
%%%%%%%%%%%%%%%%%%%%%%%%%%%%%%%%%%%
% Calibration of Interval matching
%Different offsets and minimum scores for the interval matching to trigger can be simulated and evaluated on the recorded trace until a good solution is found.\todo{insert correct values}
%This is done in a software environment as we have our ground truth available through the results of the semi-automatic locating algorithm.
We furthermore use a software environment to determine the offsets and threshold for interval matching based on the detected \ac{co} traces.

%%%%%%%%%%%%%%%%%%%%%%%%%%%%%%%%%%%
Finally, the parallel waveform-matching design \newlan{was} loaded onto the digitizer. 
With our calibrated configuration, we \newlan{found} all \emph{well-formed} encryptions. 
\newJens{
There were also some malformed \acp{co} in the side-channel trace for our experiment.
The first CO at the beginning of a batch execution or after an interrupt may well look different due to caching effects of our specific device under test and may therefore not be detected by waveform-matching.
%It is not possible to match them because they look different. 
However, this is a general challenge of waveform matching and well-known among practitioners, see~\cite{beckersDesignImplementationWaveformMatching2016}. 
%In future work, techniques based on machine learning can be applied to determine more robust templates. 
Still, techniques have been developed for successfully performing side-channel attacks on the basis of waveform matching despite such malformed \acp{co}. A basic solution here is to omit the malformed \acp{co} or treat them differently. By following such an approach, successful \ac{dpa} attacks even in the presence of not always well-formed encryptions have been shown in~\cite{balaschDPABitslicingMasking2015}.
Therefore, such attacks could beneficially use our proposed design.
%In order to also match and find differently formed COs in the side channel it might be possible to introduce an additional template $\template'$ which incorporates the special signature of those COs.
%However,  for further SCA evaluation the first few recorded COs must be omitted or treated differently in any case as they have a different length and cannot be directly compared for TVLA or CPA.  This is independent of the triggering system being used.
}

One major advantage of our measurement setup is that it does not need additional upstream analog circuitry such as an envelope detector or other filters for successful waveform matching, as is required in the related work. 
Since our waveform matching functionality is directly integrated into the pipeline of the digitizer, we can obtain traces of operations at a high sampling rate without the need for an additional oscilloscope or other hardware components.

	\section{Conclusion and Future Work}

Waveform matching is one \newlan{method of applying} \ac{sca} on real-world systems without needing to modify them.
In comparison to the state of the art, our proposed design can scale with faster ADCs and is capable of enhancing the matching speed by over $50$ times. 
This will be important to enable implementation based attacks or \ac{sca} on faster and more complex systems.

Furthermore, an implementation and evaluation of our design in the pipeline of a high-performance digitizer shows that a one-device solution is possible without sacrificing performance. 
As a consequence of the less complex measurement setup, the work of \ac{sca} evaluators in the field will also be alleviated.

One challenge that \newlan{accompanies} a parallel\newlan{, as opposed to a sequential,} design is the increased resource demands. 
However, for most examples, a template that is sub-sampled  will lower the resource demands while yielding accurate trigger results \newlan{nonetheless}, as demonstrated in our case study. 
For future work, an exploration of different techniques that \newlan{methodologically minimizes a template without compromising} accuracy is planned.
Furthermore, other matching techniques that require \newlan{fewer} features in the template $\template$ is also under current investigation.

	\newpage
	
	
	

	\bibliographystyle{ieeetr}

	%\bibliographystyle{savetrees}
	\bibliography{full_lib_at_home.bib}

\end{document}